%% file: star102309.tex
\documentclass[preprint2]{aastex}

\shorttitle{Gas Rich BTF Relation}
\shortauthors{Stark et al.}

\begin{document}

\title{A First Attempt to Calibrate the Baryonic Tully-Fisher Relation 
with Gas Dominated Galaxies}

\author{David V. Stark}
\affil{Department of Physics and Astronomy, University of North Carolina, Chapel Hill, NC 27514}

\author{Stacy S. McGaugh, Robert A. Swaters}
\affil{Department of Astronomy, University of Maryland, College Park, MD 20742}

\begin{abstract}
We calibrate the Baryonic Tully-Fisher (BTF) Relation using a sample of gas dominated galaxies. These determine the absolute scale of the baryonic mass--rotation speed relation independent of the choice of stellar mass estimator.  We find a BTF slope of $3.94\pm0.07$ (random) $\pm 0.08$ (systematic) and a zero point of $1.79\pm0.26$ (random) $\pm 0.25$ (systematic).  
We apply this relation to estimate the stellar masses of star
dominated galaxies.  This procedure reproduces the trend of mass-to-light
ratio with color predicted by population synthesis models.  The
normalization is also correct, consistent with empirical estimates of
the IMF used in such models.
\end{abstract}


\section{Introduction}

The Tully-Fisher Relation \citep{TF77} between a galaxy's luminosity and rotation velocity has been an important tool in establishing the extragalactic distance scale.  It can also be used to set an absolute scale on mass.  However, the relation between \textit{stellar mass} and rotation velocity does not hold for all galaxies \citep{MvDG,stilisrael}.  Less massive, gas rich galaxies fall below the extrapolation of the relation fit to more massive, star dominated galaxies \citep{milgbraun}.  Continuity is restored if the correlation is made between \textit{baryonic mass} and rotation velocity \citep{freeman,mcgaugh00,verh01,bdj01,gurovich,pfenniger,mcgaugh05,geha06,noordermeer,derijcke,begum08}.  This is the Baryonic Tully-Fisher (BTF) relation.

The baryonic mass of a galaxy consists of many components, including stars, dust, 
and various forms of gas.  The baryonic mass budget in spiral galaxies is
dominated by stars and cold gas: $M_b = M_s+M_g$.  By comparison, other 
known baryonic reservoirs in these systems, such as dust and ionized gas, 
contain negligible amounts of mass \citep{bregman}.

By far the greatest uncertainty is presented by the stellar mass.  While measuring
a galaxy's luminosity is straightforward, mapping $L \rightarrow M_s$ can be
fraught \citep{bdj01,piz07,AZFH}.  There are many stellar population models that provide a ratio between stellar mass and luminosity.  Unfortunately, these models can yield rather different results depending on the adopted Initial Mass Function (IMF).  The baryonic mass of a typical galaxy, with more mass in stars than in gas, is therefore uncertain.  However, the uncertainty in the baryonic mass is substantially diminished in galaxies where more of the mass is in gas than in stars.

In this paper, we make a first attempt to use gas rich galaxies to determine the BTF relationship.  We have assembled a sample with many low surface brightness and dwarf galaxies that have a higher percentage of gas than brighter high surface brightness galaxies, whose size is comparible to those originally used to calibrate the Tully-Fisher relation \citep{mdb97,sme01}.  The stellar mass is not zero, so we consider a wide range of stellar population models.  The total baryonic masses of these galaxies are determined for each model.  For each baryonic mass estimate, a BTF relation is derived.  Since these galaxies are gas dominated, the difference in the stellar mass-to-light ratio from the different population models does not have much impact, and a consistent BTF relation emerges.  

Once the BTF relation is specified by gas dominated galaxies, it can be applied to estimate the baryonic mass of star dominated galaxies.  This provides a novel estimate of their stellar masses and mass-to-light ratios.  These in turn provide an independent check on the predictions of population synthesis models.

In \S2, we describe the galaxy sample.  Its general properties are given along with a discussion of the key properties that galaxies must have to ensure a high quality sample. In \S3, the methods of finding the stellar mass are discussed. In \S4, the process of finding the BTF relation is presented and the best-fit relation is established.  In \S5, the stellar mass-to-light ratios of star dominated galaxies are determined from the BTF relation.  The results are checked against population synthesis models, and limits on the IMF are discussed.  Conclusions are given in \S6.

\section{The Sample}

Our goal is to obtain a sample of gas dominated galaxies (with $M_g > M_s$) so that an absolute calibration of the BTF relation can be made independent of stellar mass estimators.  It tends to be difficult to obtain high quality data for such objects, typically being both dwarf and low surface brightness.  Nevertheless, thanks to the efforts of many independent workers, it is now possible to assemble a sample of such galaxies that is comparable to or larger than those originally used to calibrate the Tully-Fisher relation for distance scale work.

We have scoured the literature for galaxies that meet our criteria of gas domination and well measured rotation speed.  
Our galaxy sample is primarily composed from those of McGaugh 
(2005a ---  itself a compilation of many sources) and \citet{swaters99}, with further galaxies from other sources 
\citep{deblok01,matthews08,uson03,begum08b,kassin06}.  
Such a sample can never be complete in any rigorous sense (see \S 2.4).
However, it is the saving grace of the Baryonic Tully-Fisher relation that all rotating galaxies seem to obey it.

\subsection{Data Quality}

To be included in the sample, each galaxy must satisfy certain quality criteria.
The first criterion for a galaxy to be included in the sample was that it had to have rotation curve data (line-widths are not accurate enough) and its rotation curve had to reach a constant velocity.  This flat velocity ($v_f$) is the velocity\footnote{Some workers use the maximum rather than flat velocity.  For dwarfs, these are usually identical.  For giants the difference is perceptible, but modest \citep{mcgprl,noordermeer,peskypaolo}.} used in the BTF relation.  Many apparently gas rich galaxies fail to meet this criterion.  It is common for the observed region of the rotation curve to rise continuously with no discernable flatness to be found.  This may simply be because the measurements do not extend far enough in radius to see $v_f$.  This is a common issue with dim dwarf galaxies, for which obtaining reliable data is challenging. 

\begin{figure}[h]
\epsscale{0.9}
\plotone{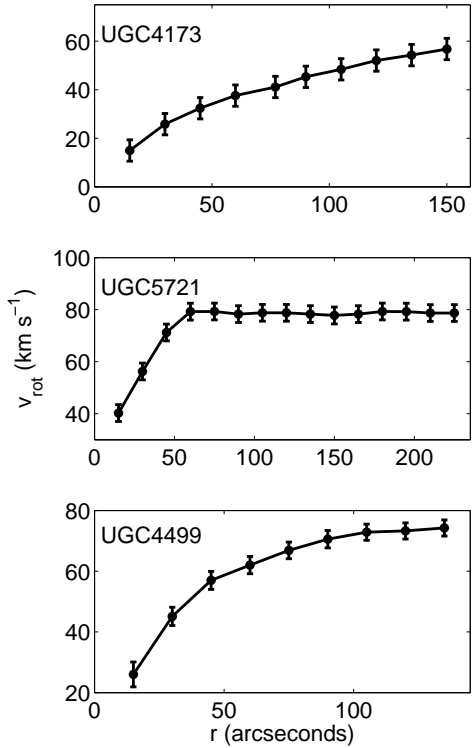}
\caption{Examples of rotation curves \citep{swaters99} 
that do and do not satisfy the flatness criterion.
The rotation curve of UGC 4173 (top) rises continuously and 
does not meet the flatness criterion.  UGC 5721 (center) is an
ideal case with clear flattening of the rotational velocity.  
UGC 4499 marginally satisfies the flatness criterion.}
\end{figure}

A rotation curve is considered to meet the flatness criterion if the measured
velocity changes by a small amount over a defined range in radius.
Quantitatively, the difference in velocity between that at three disk scale lengths and the last measured point had to be less than 15\%.  If measurements extended past four disk scale lengths, then the difference between the velocity at four disk scale lengths and the last measured point had to be less than 10\%.  Figure 1 provides examples of usable versus unusable rotation curves.

\subsection{Inclination}

We adopt the inclination given with the source data where possible.
This is usually from a tilted ring fit, but is sometimes based on
observed axis ratios.  In eight cases, the inclination was not given, so we estimated it with 
\begin{equation}
\sin{i}=\sqrt{\frac{1-\left(b/a\right)^2}{1-0.15^2}}.
\end{equation}
The values $a$ and $b$ are the semi-major and semi-minor axes of the galaxy.  We adopt 0.15 as the intrinsic thickness of a disk galaxy seen edge on \citep{kregel}.  

We select galaxies with inclinations $i >45^{\circ}$.  Since $v_f = v_{obs}/\sin{i}$, the uncertainty becomes large for smaller inclinations.  As is often the case in the literature, inclination uncertainties were not reported.  Given the adopted lower limit on the 
inclination, a reasonable and conservative estimate of the uncertainty on the inclination is $5^{\circ}$.  Choosing the minimum inclination of $45 ^{\circ}$ ensures the relative error in $v_f$ from the inclination stays below 10\%.  This way, the uncertainty in the inclination contributes to the error budget no more than most other sources of uncertainty.  Another option would have been to limit the inclination to $> 60^{\circ}$, reducing the relative error to less than 5\%.  However, this reduced the size of the galaxy sample significantly, and the number of availible galaxies is already constrained heavily by the $M_g>M_s$ requirement.  Limiting the inclination to this degree would have made the results excessively vulnerable to the inadequacies of small number statistics (see \S4.3).  As a third option, we could have allowed all galaxies with inclinations $> 30^{\circ}$, but this allowed the relative uncertainty to be as large as 30\%, making it a major contributor to the uncertainty in $v_f$.  

No upper limit on inclination was adopted.  For galaxies with very high inclinations, the 
shape of the rotation curve can be affected by internal extinction.  However, the outer,
flat portion, which is the key quantity used in the BTF relation, is usually unaffected \citep{spekkens}. 
Removing galaxies with $i >80^{\circ}$ from our sample has a neglegible effect 
on the results ($<1 \%$).  We discuss the implications of this inclination requirement in \S4.3.

\subsection{Data}

The primary criterion for inclusion in our fit to the BTF relation is gas dominance: $M_g > M_s$.  Some galaxies always satisfy this criterion while many never do. A few skirt the boundary, depending on the stellar mass estimator employed (\S 3). 

Table 1 provides basic information about the galaxies used in this study.  Both star- and gas-dominated galaxies are tabulated in order of increasing gas mass.  Though the former are not used in determining the BTF relation, we do use them to test the derived relation and stellar population models, so they are included here.  We also include several low inclination galaxies.  Again, these are not used in the determination of the BTF relation, but are used to test the implications of the $i >45^{\circ}$ limit (see \S4.3).

Many distances to galaxies were already provided in the papers in which these galaxies were found, but a search was always made for more modern or reliable distance measurements.  The most common ways the distances were determined were with a Hubble flow model ($H_0$), the red giant branch method (TRGB), the cepheid method (CEPH), and the brightest star method (BS).  Distances found with Cepheids or the red giant branch were always used if they were available, since they have been found to be more reliable indicators of a galaxy's distance \citep{karachentsev04}. Unless otherwise noted, the uncertainties for the red giant and Cepheid distance measurements were taken to be 10\%. Likewise, the uncertainty for the brightest star method was taken as 25\%.  When redshifts are used as the distance indicator,
we assume $H_0 = 75\;\textrm{km}^{-1}\,\textrm{s}^{-1}\,\textrm{Mpc}^{-1}$.
The uncertainty in this case is taken from the spread in various flow models:
\begin{equation}
\sigma_D = \frac{D_{max}-D_{min}}{2}.
\end {equation}
Here, $D_{max}$ and $D_{min}$ are the maximum and minimum estimates of the distance from the Hubble flow models provided by NED\footnote{This research has made use of the NASA/IPAC Extragalactic Database (NED) which is operated by the Jet Propulsion Laboratory, California Institute of Technology, under contract with the National Aeronautics and Space Administration.}.   It should also be noted that a number of galaxies are part of the Ursa Major cluster \citep{verh01}.  In this case, the uncertainty is that in the cluster distance \citep{tully00}.  Only a few of these galaxies are gas rich, so most do not contribute to the derivation of the BTF relation.  

The uncertainty in $v_f$  varies from case to case.  It depends on how well the flat velocity could be distinguished from the rest of the rotation curve, along with the uncertainty in the velocity measurements themselves.  The errors predominantly fell from 1-8\%, but got as high as 25\% for some cases, and 60\% for the most extreme (CamB).  

The gas mass estimates were derived from the mass of neutral hydrogen (HI), which follows the flux from the 21 cm hydrogen emission line.  We estimate the gas mass as $M_g = \frac{4}{3}M_{HI}$. The factor of 4/3 comes from the fact that hydrogen gas makes up roughly 75\% of the universe.  We ignore molecular gas, which is not measured in most of the galaxies of concern here.  In the cases where it is known \citep{bimasong}, it is always outweighed by either the stars or the atomic gas, and almost always by both.  The work of \citet{YK} suggests that this should be generally true for galaxies in our sample.  The uncertainty in the value of $M_g$ is dominated by the distance uncertainty, but there is also some error from source noise in the measurement.  

We adopt the HI masses reported by the original observers.  In most cases these are synthesis
observations (VLA or WSRT) or single dish maps (e.g., with the GMRT).  
It is common for HI fluxes to be reported without uncertainties.  
We adopt a typical error of 30\% based on the variance between the synthesis 
observations and single dish estimates. This value, coupled with the uncertainty due to distance, yields the total uncertainty in $M_g$
\begin{equation}
\sigma_{M_g}=\sqrt{\left(\frac{2\sigma_D}{D}M_g\right)^2+(0.3M_g)^2}
\end{equation}
There is no clear indication in the data we have collected that the synthesis observations
detect systematically less flux than single dish observations.

As with the gas mass, the uncertainty in luminosity is dominated by that in the distance, and again the uncertainties are seldom reported.  We adopt a 10\% uncertainty in the luminosity to reflect the likely disparity arising from many different measurements, instruments, and filters used in data collected from diverse sources in the literature.  Therefore, the total uncertainty including the affect of distance is given by
\begin{equation}
\sigma_{L}=\sqrt{\left(\frac{2\sigma_D}{D}L\right)^2+(0.1L)^2}
\end{equation}
 For the color, the $B-V$ band was used if available.  Otherwise, $B-R$ was used.  In a vast number of cases, color uncertainties are not reported.  We assume them to be neglegible.  The consequences of this assumption are discussed in \S5.2.  All sources of uncertainty are added in quadrature in the analysis.  

\subsection{Selection Effects} 

We have intentionally limited our sample to gas dominated galaxies with
reasonably high quality data.  These are predominantly, though not exclusively,
dwarfs.  Ideally, one would like a sample that is not biased towards or against any 
particular type of galaxy.  This is an impossible ideal already broken by the distinct
Tully-Fisher and Fundamental Plane relations for disk and elliptical galaxies.
Among rotating disk galaxies, typical Tully-Fisher samples are numerically 
dominated by bright galaxies with $v_f > 100 \textrm{km}\,\textrm{s}^{-1}$ 
simply because they are easier to see.  This type of bias is unavoidable.

Perhaps the most remarkable aspect of the BTF relation is to the extent to
which galaxies of very different types appear to obey it.  The relation is 
continuous over five decades in stellar mass \citep{mcgaugh05}, 
from the fastest rotators 
($\sim 300\;\textrm{km}\,\textrm{s}^{-1}$) to the slowest 
($\sim 20\;\textrm{km}\,\textrm{s}^{-1}$).  There is no persuasive evidence
for a second parameter in the relation, as might be expected for surface
brightness \citep{aaronson,zwaan,mdb98} 
or scale length \citep{courtrix,mcgprl}.
If disk galaxies do 
indeed all fall on a single BTF relation, as appears to be the case, then it hardly
matters what portion of the relation we sample in order to calibrate it.  We do of
course wish to sample as broadly in velocity as possible in order to
constrain the slope.  Nonetheless, if the relation is both valid and universal,
calibrating it with lower velocity gas rich dwarfs is no worse than calibrating
it with high velocity star dominated giants as is usually done.  Indeed, it has
the advantage of providing an absolute normalization to the mass scale.  

\section{Finding the Stellar and Baryonic Mass}

The true stellar mass of galaxies is unknown.
We estimate $M_s$ with a variety of stellar population models. 
These models take the general form
\begin{equation}
\log{\frac{M_s}{L}}=\log{\Upsilon}=q_B+\left(s_B\times color\right).
\end{equation}
The constants $q_B$ and $s_B$ depend on the IMF and the luminosity band.  
We use $B$-band luminosities throughout here, as this is what is most commonly available.  Though not as desirable as the $K$-band for purposes of stellar
mass estimates, this is not a substantial issue here given the selection for gas dominated galaxies.  Which galaxies are determined to be gas dominated does depend on the IMF.  

We utilize the stellar population synthesis models of \citet{Bell03} and
\citet{portinari04}.  These workers find that the slope $s_B$ is fairly insensitive
to the choice of IMF, the chief effect being on the normalization $q_B$.  
We used the Kroupa, Salpeter, and Kennicutt IMF models from \citet{portinari04}, and the Scaled Salpeter, Kroupa, and Bottema IMF models from \citet{Bell03}.  It should be noted that the two models give different mass-to-light ratios for what is nominally the same IMF.  This appears to stem from differing treatments of brown dwarfs as well as other differences in the modeling.  Nevertheless, the models are in fairly good agreement.

The population models are only expected to be valid over a finite range of color
\citep{portinari04}.  Some galaxies have such blue colors that they fall outside of this range.
For lack of a better approach, we assume equation (5) holds for all colors.  
This does not impact the baryonic mass of gas dominated galaxies since blue colors 
imply low mass-to-light ratios regardless of the accuracy of equation (5). 
It hardly matters if $\Upsilon$ is tiny or merely small.  

The stellar mass is calculated with
\begin{equation}
M_s = {\Upsilon}L.
\end{equation}
The uncertainty in the mass-to-light ratio for Portinari's and Bell's population models is based on their estimates of the inherent scatter in their relations.  These are 0.1 and 0.15 dex for Bell's and Portinari's models, respectively.  

The total baryonic mass is simply the sum of the galaxy's gaseous and stellar components
\begin{equation}
M_b=M_g+M_s.
\end{equation}
For all of these calculations, the uncertainties are added in quadrature.  Tables 2 and 3 show the results from Portinari's and Bell's models. For each galaxy in the sample, the mass-to-light ratio ($\Upsilon$), stellar mass ($M_s$), and total baryonic mass ($M_b$) are provided, as given by each stellar population model.  We also note which galaxies qualify as gas dominated, with
$M_g > M_s$.  These are the galaxies we use to calibrate the BTF relation.  Galaxies that are star dominated under all population models are not included.

\section{Determining the BTF Relation}

\subsection{Gas Dominated Subsamples}

We have made six separate stellar mass estimates per galaxy.  These are hereafter named for the population model-IMF combination utilized (Table 4).  The idea is to only use the gas dominated galaxies to determine the BTF relation.  For each sub-sample, only galaxies with gas mass greater than stellar mass 
($M_g > M_s$: Tables 2 and 3) are used to make fits. 

The number of galaxies in each sub-sample differs because of the different stellar mass estimators.  Under one population model, a galaxy might have more gas mass than stellar mass, while under another model, the opposite could be the case.  The final number of galaxies in each sub-sample is given in Table 4.  

For comparison, the median mass-to-light ratios (in $M_{\sun}/L_{\sun}$)
of the sub-samples are 0.61 for Portinari-Kroupa, 
0.87 for Portinari-Salpeter, 0.51 for Portinari-Kennicutt,
0.67 for Bell-Scaled Salpeter, 0.47 for Bell-Kroupa, and 0.28 for Bell-Bottema.  Note that in all cases, the typical mass-to-light ratio is less than unity.  
This is perhaps not surprising in that galaxies are more likely
to be found to be gas dominated if they have low $\Upsilon$.  However, we also note that some prescriptions seem to 
give implausibly low mass-to-light ratios.  This is particularly
true of the Bell-Bottema case, which leads to the inclusion of many galaxies that
would not otherwise be considered gas dominated (e.g., NGC 2998).

\subsection{Linear Fits to the Subsamples}

We assume the BTF relation is linear in the logarithm.  
If there is any curvature in the intrinsic relation, it is not apparent in our data.  Therefore, fits of the form
\begin{equation}
\log{M_b}=x\log{v_f}+A
\end{equation}
were made to each of the six gas-dominated subsamples.  We used the OLS bisector linear fit method to determine the BTF relation coefficients. 
Forward and reverse fits were first conducted following the standard $\chi^2$ minimization routine outlined in \citet[chap. 15.3]{Press}, taking into account uncertainties in both $M_b$ and $v_f$.  Using the coefficients from these fits and their respective uncertainties, the final OLS bisector slope and intercept were determined following the technique in \cite{isobe90}.   

The hope and intent was that the choice of stellar population model would have a little effect on the total baryonic mass, and in turn, on the derived BTF relation.  Table 4 shows the results of the 
fit for each gas rich sub-sample.  One can immediately see that indeed the population model 
does not have a drastic effect on the BTF relation.  

As a further test, we considered the selection criterion $M_g > 2M_s$ to further suppress any dependence on the choice of population model.  The results are
indistinguishable.  Rather few galaxies survive this very restrictive selection 
criterion, so we do not consider this case further.

The choice of stellar mass estimator has an impact on the quality of the
fits.  Three cases (the Portinari-Kroupa, Portinari-Salpeter, and Bell-Scaled Salpeter 
sub-samples) have reasonable reduced $\chi^2$ values (Table 4).  These are 
shown in Figure 2.  The other cases look similar, albeit with worse formal $\chi^2$.
In order to determine a single calibration for the BTF relation, a weighted average is made of the coefficients from the three illustrated sub-samples, with weights given by
\begin{equation}
w=\frac{1}{\mid \chi^2-1 \mid }.
\end{equation}
The resulting BTF relation is
\begin{equation}
\log M_b = 3.94 \log v_f + 1.79.
\end{equation}
The formal random uncertainty in this OLS bisector fit is $\pm 0.07$ in the slope and $\pm 0.26$
in the intercept.  
In addition to the random errors, we attempt to estimate the systematic uncertainties.  The choice of population model still plays a some role in constraining the intercept of the BTF relation, as the stellar mass is minimized but not zero. Likewise, there is some systematic uncertainty in the slope that seems to depend
on the inclination cut off.  This will be dicussed in further detail in \S4.3.  For now, the estimated magnitude
of the systematic uncertainties are $\pm 0.08$ in the slope and $\pm 0.25$ in the intercept.  

\begin{figure}[h]
\epsscale{0.9}
\plotone{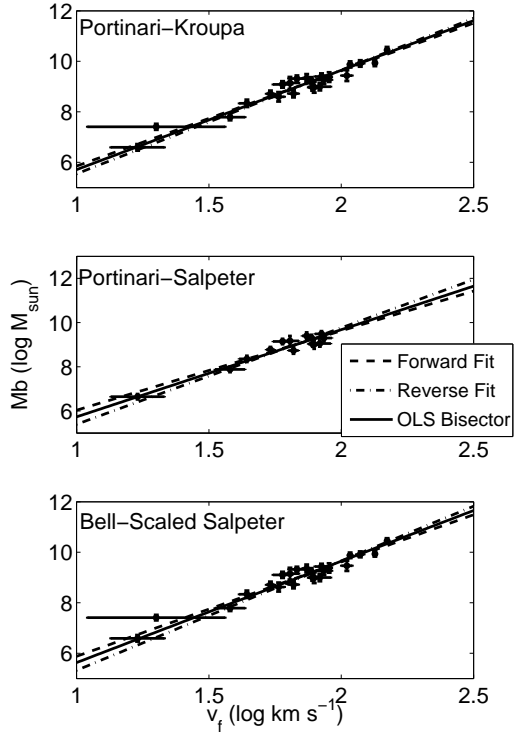}
\caption{Fits of the BTF relation to the three gas dominated sub-samples
with good $\chi_{\nu}^2$.}
\end{figure}

\subsection{Systematic Effects and the Inclination Limit}

We have investigated how our sample selection criteria affect the result.
The choice of the stellar mass estimator, the line between star and gas domination,
and the velocity measure employed ($v_f$ vs.\ $v_{max}$)
all make little difference.  The limit on inclination, $i > 45^{\circ}$, has a mild effect on the
slope, discussed further below.  While variations in the sample selection makes no net
systematic difference to the intercept, the range of variation does gives some handle on
the possible amplitude of the systematic error.

The formal error in the intercept has been determined by bootstrap resampling of the data.  
Approximately 95\% of the data fall within 2$\sigma$ of the best fit value.  As is usually the case
for astronomical data, it is rather harder to estimate the systematic uncertainty.  By examining the
variation in the zero points from the fits to the various subsamples, forward and reverse 
OLS as well as the bisector method, we estimate a systematic uncertainty in the intercept 
of $\pm 0.25$.  This covers the full range of variation, so is rather conservative.  

For the uncertainty in slope, there does seem to be a mild but systematic dependence on the
imposed inclination limit.  Some cut off on the inclination of galaxies included in the sample is
necessary, as the velocity depends on $1/sin(i)$.  Face on galaxies have very large
uncertainties due to inclination.  Inclusion of such data will bias the determination of the slope
\citet{Jefferys80}, so we impose a limit to combat this effect.
This inclination limit is chosen to provide a middle ground between data quality and sample size (see \S2.2).  As a check, we vary the inclination limit imposed on the gas dominated sub-samples, letting the minimum range from 0 to $70^{\circ}$.  The BTF relation is derived for each inclination cut.  The bisector slope and sample size of each fit is shown in Figure 3.  

\begin{figure}[h]
\epsscale{2.2}
\plottwo{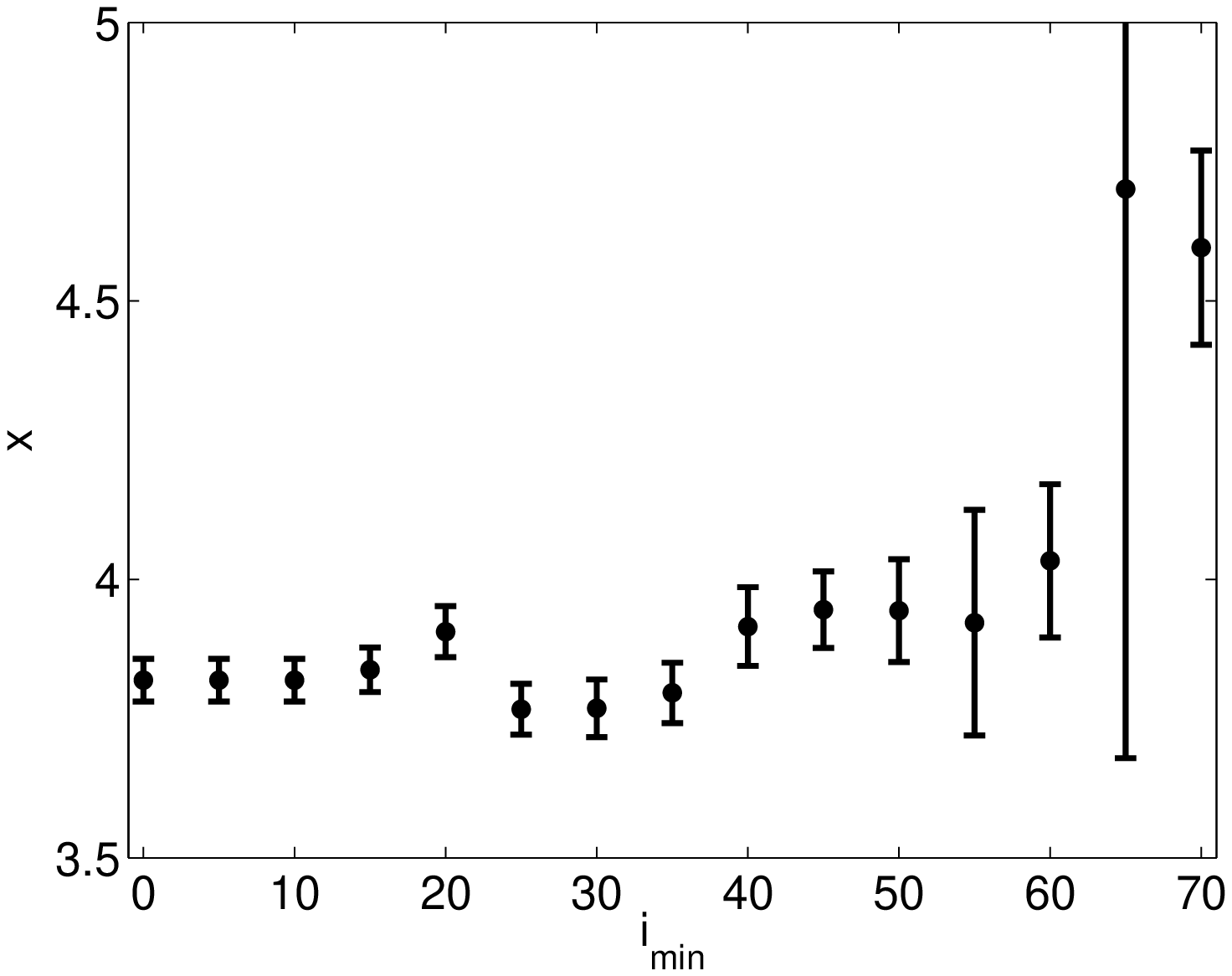}{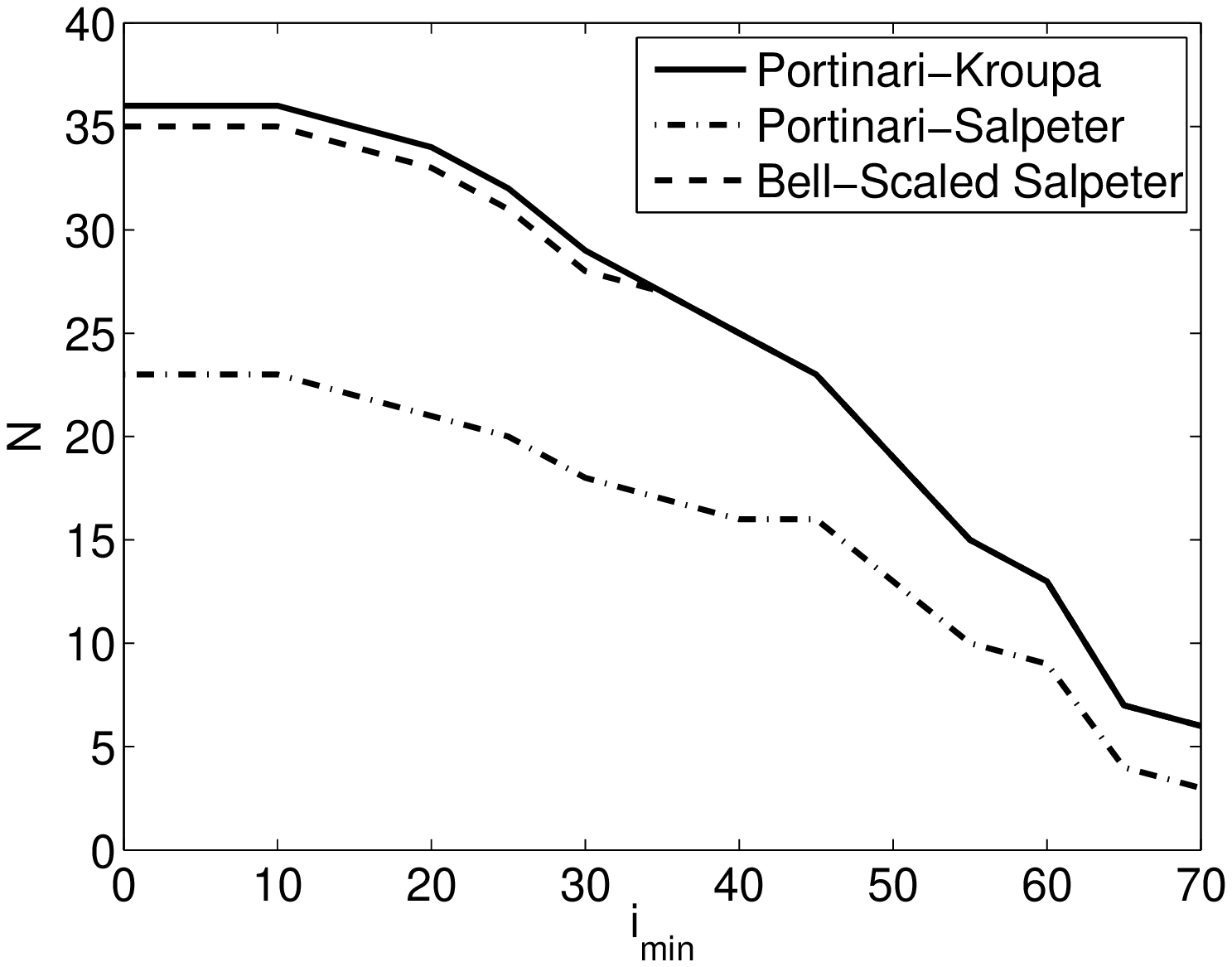}
\caption{Derived slope ($x$) of the BTF fit (left) and number of galaxies (right) versus the minimum inclination limit imposed upon the sample.}
\end{figure}

The slope is relatively stable from no cut to an inclination limit of $i > 55^{\circ}$, varying between 3.76 and 3.95.  For the most part, the magnitude of these variations are comparable to the uncertainty in the fits.  
For higher inclination cuts, the slope becomes much larger.  The cause of this is the drastic reduction in sample size, as seen in figure 4.  For the last inclination cut, only 15\% of the sample remains, leaving only 3 to 6 galaxies depending on the population model used.  Therefore, we reach a point where restricting the inclination too much simply leads to small number statistics.  This effect sets in around $i_{min}\approx60^{\circ}$, where the slope strongly deviates from the mean, and the uncertainty drastically increases.  This is most clearly seen in the transition from $i_{min}=60^{\circ}$ to $i_{min}=65^{\circ}$.  Over this range, the sample size is effectively cut in half.  This leaves one galaxy with a low relative velocity, but it has a very large uncertainty, effectively leaving nothing to constrain the low end of the fit.  Choosing $i > 45^{\circ}$ keeps us well away from this region where small number statistics issues like this one take over.  Accounting for the variation in best fit slope with the choice of inclination
limit gives a systematic uncertainty of 0.08, discounting the sharp change due to small numbers at 
very large inclination limits.



\section{Stellar Masses from the BTF Relation }

The Baryonic Tully-Fisher Relation derived from gas dominated galaxies provides a novel estimator of a galaxy's total baryonic mass.  The observed rotation speed 
$v_f$ is in effect a measure of baryonic mass.  For a galaxy of known distance,
the stellar mass can be found by subtracting off the observed gas mass:
\begin{equation}
M_s=M_b-M_g.
\end{equation}
Dividing this by the luminosity gives the mass-to-light ratio, $\Upsilon$.

Table 5 shows the baryonic mass, stellar mass, and mass-to-light ratio indicated 
by the BTF relation.  Note that this relation, once derived, can be applied to any
rotating galaxy.  This provides an estimate of the stellar mass for the star dominated
galaxies in Table 1 that is independent of population synthesis models.

\subsection{Testing Population Synthesis Models}

The stellar mass-to-light ratios determined from the BTF relation provide a means of testing the population synthesis models.  These models utilize our knowledge of stellar evolution 
and prescriptions for galactic star formation histories to predict the 
color-$\Upsilon$ relation.  We utilize dynamical information
in the form of the BTF relation to
estimate the mass-to-light ratios of star dominated galaxies.  

The methods and data involved in this process are distinct and very nearly independent.
They are not completely independent as population synthesis models have been used in 
the BTF calibration.  However, as revealed by Table 4, the differences in stellar mass given by the different population models have only a neglegible impact on the BTF.
By construction, the gas mass dominates the calibration.
Consequently, this may be as close as it is possible to get to an independent 
test of the predictions of population synthesis models.

We detect a clear correlation between mass-to-light ratio and color.
This is expected in the models of
\citet{portinari04} and \citet{Bell03}, which are consistent in predicting a 
slope of $s_B = 1.7$ for the color-$\Upsilon$ relation.  
Our data are consistent with this slope (Figure 4).  
This appears to be a remarkable confirmation of 
a basic prediction of population synthesis models.

\subsection{Constraining the IMF}

The main uncertainty in population models is from the IMF.  Here
we apply our calibration of the BTF relation to derive our own color-$\Upsilon$ relation. 
This provides an independent constraint on the IMF.

We fit the IMF normalization $q_B$ of equation (5) to the data in Figure 4.
The slope of the data is consistent with the expectation of the models, within the 
uncertainties.  To provide a direct comparison to the IMF of the models, we
fix the slope and fit only for the normalization $q_B$.
In this process, we utilize only those star dominated galaxies with $B-V$ colors that 
were not part of the three sub-samples used to calibrate the BTF relation (\S2.4).

The BTF-derived IMF coefficient provides a constraint on the IMF.  
We find $q_B = -0.94\pm 0.20$ such that, in the mean,
\begin{equation}
\log{\Upsilon}=1.7\left(\bv\right)-0.94.
\end{equation}
The BTF-derived values of $\Upsilon$ and the fit (equation 12) 
are shown in Figure 4.  

\begin{figure}[h]
\epsscale{.95}
\plotone{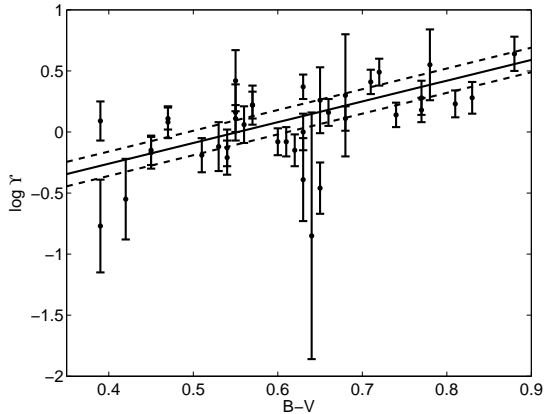}
\caption{The mass-to-light ratios $\Upsilon$ derived from the BTF relation as a function of $B-V$ color.  The line represents a fit to these data.  Only star dominated galaxies are used, not the gas dominated galaxies used to determine the BTF.  The normalization is the only fitting parameter; the slope is fixed to the value predicted by stellar population models \citep{Bell03, portinari04}.  
The dotted lines show the uncertainty in the fit.}
\end{figure}

The constraint derived here on the IMF agrees most closely with the Bell-Scaled Salpeter or Portinari-Kroupa models.  
This is not a strong statement, as the uncertainties 
allow many of the IMFs considered by both \citet{Bell03} and \citet{portinari04}.
However, it does suggest that the IMF is normal, and not outrageously different
than assumed in population synthesis models.  Indeed, the more extreme IMFs,
both heavy (Portinari-Salpeter) and light (Bell-Bottema), are disfavored.

The scatter in $\Upsilon$ is 0.20 dex.  Roughly half of this can be attributed to observational uncertainties,
leaving an intrinsic scatter of $\sim 0.14$ in the mass-to-light ratio.  This is consistent with the intrinsic 
scatter expected by \citet{Bell03} and \citet{portinari04}.  
	
As mentioned in \S2.3, we take the color uncertainties to be neglegible as they are widely unreported.  We test the implication of this assumption by repeating the linear fits to find $q_B$ but giving the $B-V$ values uniform uncertainties.  The IMF normalization $q_B$ goes down by 0.02 for $\sigma_{B-V} = 
0.05$, then drops by 0.04 for every addition of 0.05 to the color 
uncertainty up to $\sigma_{B-V} = 0.2$, the largest value considered.  It seems unlikely that the average uncertainty in the colors would be larger than this, so the effect is likely to be modest.
However, we can not exclude the possibility of a larger effect if the galaxies with extremal colors also have large uncertainties.  
 
\section{Conclusions}

We have fit the Baryonic Tully-Fisher relation for a sample of gas dominated
galaxies.  We obtain an absolute calibration of the baryonic mass-rotation 
speed relation that is effectively independent of the choice of stellar mass
estimator.  For the data assembled here, we find a BTF relation with a slope
$x = 3.94 \pm 0.07$ (random) $\pm 0.08$ (systematic) and an intercept $A = 1.79\pm0.26$
(random) $\pm 0.25$ (systematic).

We use our derived BTF relation to estimate the baryonic and stellar masses
of a sample of star dominated galaxies independent of the calibration sample.
The stellar mass-to-light ratios found in this way are in remarkably good
agreement with the predictions of population synthesis models
\citep{Bell03,portinari04}.  The expected trend of mass-to-light ratio
with color is reproduced, and the absolute scale is consistent with the
best estimates of the IMF.  

To place a constraint on the IMF, we have treated the coefficient $q_B$ 
of the color--mass-to-light ratio relation (equation 5) as a fit parameter.  
We obtain $q_B = -0.94 \pm 0.20$.
This is consistent with a Kroupa, Kennicutt, or Scaled Salpeter IMF.
When observational uncertainty is subtracted, the amount of scatter 
in the color-mass-to-light ratio relation is consistent with that of 
past models, and does not
exceed that expected from variation in the star formation histories of
different galaxies.  This suggests that variation in the IMF does not contribute
significantly to the scatter.  Perhaps the IMF tends to some universal
value when averaged over the many individual star formation events
necessary to build up the stellar mass of bright galaxies.

A common interpretation of the Tully-Fisher relation is that baryonic mass
scales with dark mass, and the observed velocity is proportional to the
characteristic halo velocity.  These assumptions can not strictly hold, as
the observed slope deviates significantly from that ($x=3$)
of the theoretical halo mass-velocity relation for CDM halos \citep{steinmetz99}.
Perhaps the effective baryon fraction varies systematically
with circular velocity, or some other physics underlies the relation.

This study was a first attempt to calibrate the BTF relation
using gas dominated galaxies.  However, in its current state, it is not free from
potential statistical uncertainties (\S4.3).  This is quite evident by our measure of $\chi_{\nu}^2$ as a function of inclination limit.  Even a small adjustment to the sample can change the fit, though for the most part we consider this to be within tolerable limits.  Future attempts to calibrate the BTF relation would benefit from a 
larger sample of gas rich galaxies.

\acknowledgements 
The work of SSM is supported in part by NSF grant AST 0505956.

\clearpage






\clearpage

\input{t1}

\input{t2}

\input{t3_corr}

\clearpage

\input{t4_corr}

\input{t5_corr}

\clearpage

\end{document}

%% file: t1.tex
\begin{deluxetable}{lrlccccccccc}

\tabletypesize{\footnotesize}
\tablewidth{0pt}
\tablecaption{Galaxy Sample}
\tablehead{
\colhead{Galaxy}&\colhead{$D^a$}&\colhead{$\sigma_D$} &\colhead{Method}&\colhead{Ref.}&\colhead{$i$}&\colhead{$v_f$}&\colhead{$\sigma_{vf}$}&\colhead{$M_g$}&\colhead{$L_B$}&\colhead{Color}&\colhead{Band}\\
\colhead{} & \multicolumn{2}{c}{(Mpc)} & \colhead{} & \colhead{} & \colhead{(degrees)}&\multicolumn{2}{c}{($\textrm{km}\,\textrm{s}^{-1}$)} & \colhead{($\log{M_{\sun}}$)} & \colhead{($\log{L_{\sun}}$)} & \colhead{} & \colhead{}\\
\colhead{(1)} & \colhead{(2)} & \colhead{(3)} & \colhead{(4)} & \colhead{(5)} & \colhead{(6)} & \colhead{(7)} & \colhead{(8)} & \colhead{(9)} & \colhead{(10)} & \colhead{(11)} & \colhead{(12)}}
\startdata

DDO210	&	\phn0.94	&	0.09	&	TRGB	&	\phn9	&	$50^b$	&	\phn17	&	\phn4	&	\phn6.51	&	\phn6.38	&	0.25	&	$B-V$	\\
CamB	&	\phn3.34	&	0.33	&	TRGB	&	\phn4	&	65	&	\phn20	&	12	&	\phn7.22	&	\phn6.93	&	0.58	&	$B-V$	\\
WLM	&	\phn0.95	&	0.1\phn	&	TRGB	&	10	&	64	&	\phn38	&	\phn5	&	\phn7.60	&	\phn7.77	&	0.29	&	$B-V$	\\
KK98251	&	\phn5.9\phn	&	1.48	&	BS	&	\phn9	&	62	&	\phn36	&	\phn6	&	\phn8.06	&	\phn8.05	&	0.79	&	$B-V$	\\
NGC3741	&	\phn3.0\phn	&	0.3\phn	&	TRGB	&	\phn8	&	$46^b$	&	\phn44	&	\phn3	&	\phn8.3	&	\phn7.52	&	0.38	&	$B-V$	\\
UGC8550	&	\phn5.1\phn	&	1.28	&	BS	&	\phn9	&	90	&	\phn58	&	\phn3	&	\phn8.34	&	\phn7.88	&	1.26	&	$B-R$	\\
DDO168	&	\phn4.33	&	0.43	&	TRGB	&	\phn4	&	$60^b$	&	\phn54	&	\phn2	&	\phn8.60	&	\phn8.46	&	0.32	&	$B-V$	\\
NGC3109	&	\phn1.33	&	0.13	&	TRGB	&	\phn2	&	86	&	\phn66	&	\phn3	&	\phn8.70	&	\phn7.54	&	0.47	&	$B-V$	\\
UGC8490	&	\phn4.65	&	0.47	&	TRGB	&	\phn9	&	50	&	\phn78	&	\phn3	&	\phn8.84	&	\phn8.91	&	0.48	&	$B-R$	\\
UGC3711	&	\phn7.9\phn	&	0.79	&	TRGB	&	\phn9	&	60	&	\phn95	&	\phn3	&	\phn8.89	&	\phn8.80	&	1.05	&	$B-R$	\\
F565V2	&	48\phd		&	6.65	&	H0	&	\phn3	&	60	&	\phn83	&	\phn8	&	\phn8.91	&	\phn8.36	&	0.51	&	$B-V$	\\
UGC5721	&	\phn6.5\phn	&	1.63	&	AVG	&	\phn9	&	61	&	\phn79	&	\phn3	&	\phn8.93	&	\phn8.55	&	0.63	&	$B-R$	\\
UGC3851	&	\phn3.19	&	0.32	&	TRGB	&	\phn9	&	59	&	\phn60	&	\phn5	&	\phn8.97	&	\phn8.53	&	0.88	&	$B-R$	\\
NGC2683	&	\phn7.73	&	2.32	&	SBF	&	\phn1	&	82	&	155	&	\phn5	&	\phn9.06	&	10.14	&	0.65	&	$B-V$	\\
UGC6667	&	18.6\phn	&	1.93	&	GROUP 	&	\phn5	&	90	&	\phn85	&	\phn3	&	\phn9.06	&	\phn9.57	&	0.65	&	$B-V$	\\
UGC6923	&	18.6\phn	&	1.93	&	GROUP 	&	\phn5	&	65	&	\phn81	&	\phn5	&	\phn9.06	&	\phn9.50	&	0.42	&	$B-V$	\\
IC2574	&	\phn4.02	&	0.4\phn	&	TRGB	&	\phn4	&	$58^b$	&	\phn68	&	\phn5	&	\phn9.08	&	\phn9.16	&	0.42	&	$B-V$	\\
UGC9211	&	12.6\phn	&	3.3\phn	&	GROUP	&	11	&	48	&	\phn64	&	\phn5	&	\phn9.12	&	\phn8.39	&	0.72	&	$B-R$	\\
NGC1560	&	\phn3.45	&	0.35	&	TRGB	&	\phn2	&	80	&	\phn77	&	\phn1	&	\phn9.10	&	\phn8.67	&	0.57	&	$B-V$	\\
NGC5585	&	\phn5.7\phn	&	1.45	&	BS	&	\phn1	&	51	&	\phn90	&	\phn2	&	\phn9.15	&	\phn9.13	&	0.46	&	$B-V$	\\
UGC7321	&	10\phd\phn\phn	&	3\phd\phn\phn &	AVG	&	12	&	90	&	105	&	\phn5	&	\phn9.20	&	\phn9.03	&	0.97	&	$B-R$	\\
UGC6818	&	18.6\phn	&	1.93	&	GROUP 	&	\phn5	&	$68^b$	&	\phn72	&	\phn6	&	\phn9.16	&	\phn9.41	&	0.43	&	$B-V$	\\
NGC7793	&	\phn3.91	&	0.39	&	TRGB	&	\phn1	&	47	&	\phn95	&	10	&	\phn9.20	&	\phn9.73	&	0.63	&	$B-V$	\\
UGC4499	&	13\phd\phn\phn	&	2.35	&	H0	&	11	&	50	&	\phn74	&	\phn3	&	\phn9.20	&	\phn9.02	&	0.72	&	$B-R$	\\
IC 2233	&	10.4\phn	&	1.04	&	TRGB	&	\phn7	&	89	&	\phn84	&	\phn5	&	\phn9.20	&	\phn9.29	&	0.67	&	$B-R$	\\
NGC3972	&	18.6\phn	&	1.93	&	GROUP 	&	\phn5	&	77	&	133	&	\phn3	&	\phn9.24	&	\phn9.99	&	0.55	&	$B-V$	\\
F583-1	&	32\phd\phn\phn	&	5.25	&	H0	&	13	&	63	&	\phn86	&	\phn6	&	\phn9.20	&	\phn8.67	&	0.39	&	$B-V$	\\
NGC4085	&	18.6\phn	&	1.93	&	GROUP 	&	\phn5	&	82	&	136	&	\phn6	&	\phn9.27	&	10.07	&	0.47	&	$B-V$	\\
NGC6503	&	\phn5.27	&	0.53	&	TRGB	&	\phn2	&	74	&	115	&	\phn1	&	\phn9.28	&	\phn9.58	&	0.57	&	$B-V$	\\
NGC3877	&	18.6\phn	&	1.93	&	GROUP 	&	\phn5	&	76	&	170	&	\phn1	&	\phn9.30	&	10.45	&	0.68	&	$B-V$	\\
NGC4138	&	18.6\phn	&	1.93	&	GROUP 	&	\phn5	&	53	&	148	&	\phn4	&	\phn9.30	&	10.07	&	0.81	&	$B-V$	\\
NGC247	&	\phn3.65	&	0.37	&	TRGB	&	\phn1	&	74	&	106	&	\phn2	&	\phn9.34	&	\phn9.77	&	0.54	&	$B-V$	\\
UGC6973	&	18.6\phn	&	1.93	&	GROUP 	&	\phn5	&	71	&	174	&	\phn9	&	\phn9.39	&	\phn9.95	&	0.88	&	$B-V$	\\
NGC3917	&	18.6\phn	&	1.93	&	GROUP 	&	\phn5	&	79	&	137	&	\phn1	&	\phn9.41	&	10.21	&	0.6	&	$B-V$	\\
UGC6917	&	18.6\phn	&	1.93	&	GROUP 	&	\phn5	&	57	&	111	&	\phn5	&	\phn9.46	&	\phn9.74	&	0.53	&	$B-V$	\\
NGC4217	&	18.6\phn	&	1.93	&	GROUP 	&	\phn5	&	86	&	178	&	\phn1	&	\phn9.56	&	10.44	&	0.77	&	$B-V$	\\
NGC4051	&	18.6\phn	&	1.93	&	GROUP 	&	\phn5	&	49	&	160	&	\phn7	&	\phn9.57	&	10.57	&	0.62	&	$B-V$	\\
NGC3953	&	18.6\phn	&	1.93	&	GROUP 	&	\phn5	&	62	&	223	&	\phn3	&	\phn9.59	&	10.62	&	0.71	&	$B-V$	\\
NGC4010	&	18.6\phn	&	1.93	&	GROUP 	&	\phn5	&	$76^b$	&	123	&	\phn3	&	\phn9.59	&	\phn9.96	&	0.54	&	$B-V$	\\
NGC4013	&	18.6\phn	&	1.93	&	GROUP 	&	\phn5	&	$80^b$	&	177	&	\phn7	&	\phn9.62	&	10.32	&	0.83	&	$B-V$	\\
UGC6983	&	18.6\phn	&	1.93	&	GROUP 	&	\phn5	&	49	&	108	&	\phn2	&	\phn9.62	&	\phn9.69	&	0.45	&	$B-V$	\\
NGC4100	&	18.6\phn	&	1.93	&	GROUP 	&	\phn5	&	73	&	160	&	\phn9	&	\phn9.64	&	10.41	&	0.63	&	$B-V$	\\
NGC2403	&	\phn3.18	&	0.32	&	CEPH	&	\phn2	&	60	&	134	&	\phn1	&	\phn9.65	&	\phn9.88	&	0.39	&	$B-V$	\\
NGC3949	&	18.6\phn	&	1.93	&	GROUP 	&	\phn5	&	55	&	165	&	10	&	\phn9.68	&	10.38	&	0.39	&	$B-V$	\\
NGC4183	&	18.6\phn	&	1.93	&	GROUP 	&	\phn5	&	82	&	111	&	\phn2	&	\phn9.69	&	10.11	&	0.39	&	$B-V$	\\
F5741	&	96\phd		&	6.3\phn	&	H0	&	\phn3	&	$82^b$	&	\phn99	&	\phn1	&	\phn9.69	&	\phn9.57	&	1.06	&	$B-R$	\\
F5681	&	85\phd		&	10.1	&	H0	&	\phn3	&	46	&	118	&	\phn4	&	\phn9.70	&	\phn9.45	&	0.58	&	$B-V$	\\
NGC2903	&	\phn8.9\phn	&	2.23	&	BS	&	\phn2	&	63	&	181	&	\phn4	&	\phn9.78	&	10.47	&	0.55	&	$B-V$	\\
NGC3521	&	\phn9.3\phn	&	5\phd\phn\phn &	H0	&	\phn1	&	61	&	190	&	15	&	\phn9.84	&	10.42	&	0.68	&	$B-V$	\\
NGC3769	&	18.6\phn	&	1.93	&	GROUP 	&	\phn5	&	70	&	118	&	\phn5	&	\phn9.88	&	\phn9.99	&	0.64	&	$B-V$	\\
NGC3893	&	18.6\phn	&	1.93	&	GROUP 	&	\phn5	&	49	&	176	&	\phn9	&	\phn9.91	&	10.49	&	0.56	&	$B-V$	\\
NGC1003	&	12.3\phn	&	3.41	&	H0	&	\phn1	&	67	&	112	&	\phn3	&	\phn9.95	&	10.21	&	0.55	&	$B-V$	\\
NGC3726	&	18.6\phn	&	1.93	&	GROUP 	&	\phn6	&	53	&	168	&	\phn2	&	\phn9.95	&	10.58	&	0.45	&	$B-V$	\\
NGC 157	&	20.16		&	4.65	&	H0	&	\phn7	&	45	&	120	&	17	&	\phn9.99	&	10.34	&	0.62	&	$B-V$	\\
NGC5033	&	12.4\phn	&	3.35	&	H0	&	\phn1	&	64	&	195	&	\phn4	&	10.00	&	10.31	&	0.55	&	$B-V$	\\
NGC5907	&	11.0\phn	&	3.58	&	H0	&	\phn1	&	90	&	215	&	\phn2	&	10.04	&	10.38	&	0.78	&	$B-V$	\\
NGC7331	&	15.1\phn	&	1.51	&	CEPH	&	\phn2	&	75	&	238	&	\phn2	&	10.05	&	10.74	&	0.63	&	$B-V$	\\
NGC4088	&	18.6\phn	&	1.93	&	GROUP 	&	\phn5	&	69	&	170	&	\phn5	&	10.06	&	10.61	&	0.51	&	$B-V$	\\
NGC4157	&	18.6\phn	&	1.93	&	GROUP 	&	\phn5	&	82	&	185	&	\phn1	&	10.06	&	10.46	&	0.66	&	$B-V$	\\
NGC3992	&	18.6\phn	&	1.93	&	GROUP 	&	\phn5	&	56	&	241	&	\phn4	&	10.12	&	10.65	&	0.72	&	$B-V$	\\
NGC3198	&	14.45		&	1.45	&	CEPH	&	\phn2	&	74	&	149	&	\phn1	&	10.17	&	10.32	&	0.43	&	$B-V$	\\
NGC2998	&	67\phd		&	7\phd\phn\phn &	H0	&	\phn1	&	63	&	212	&	\phn3	&	10.48	&	10.95	&	0.45	&	$B-V$	\\
NGC801	&	83\phd		&	8.1\phn	&	H0	&	\phn1	&	80	&	216	&	\phn1	&	10.49	&	10.90	&	0.61	&	$B-V$	\\
NGC5533	&	56\phd		&	7.05	&	H0	&	\phn1	&	55	&	240	&	\phn6	&	10.51	&	10.78	&	0.77	&	$B-V$	\\
NGC2841	&	14.1\phn	&	1.41	&	CEPH	&	\phn2	&	70	&	287	&	\phn4	&	10.58	&	11.28	&	0.74	&	$B-V$	\\
NGC6674	&	49\phd		&	7.5\phn	&	H0	&	\phn1	&	46	&	242	&	\phn2	&	10.59	&	10.83	&	0.57	&	$B-V$	\\
UGC2885	&	82\phd		&	6.95	&	H0	&	\phn1	&	64	&	298	&	\phn1	&	10.73	&	11.35	&	0.47	&	$B-V$	\\
\tableline
GR8	&	\phn2.1\phn	&	0.21	&	TRGB	&	14	&	28	&	\phn25	&	\phn5	&	\phn7.10	&	\phn6.99	&	0.38	&	$B-V$	\\
DDO154	&	\phn4.3\phn	&	1.08	&	BS	&	\phn3	&	28	&	\phn48	&	\phn6	&	\phn8.72	&	\phn7.76	&	0.32	&	$B-V$	\\
UGC7278	&	\phn2.94	&	0.29	&	TRGB	&	\phn9	&	30	&	\phn80	&	\phn2	&	\phn8.88	&	\phn9.05	&	0.8	&	$B-R$	\\
UGC5846	&	13.2\phn	&	3.55	&	GROUP	&	11	&	30	&	\phn51	&	\phn7	&	\phn8.97	&	\phn8.47	&	0.39	&	$B-R$	\\
UGC4325	&	10.1\phn	&	1.85	&	H0	&	11	&	41	&	\phn92	&	\phn2	&	\phn9.00	&	\phn9.16	&	0.68	&	$B-R$	\\
E215g9	&	\phn5.25	&	0.41	&	TRGB	&	15	&	36	&	\phn51	&	\phn5	&	\phn9.03	&	\phn7.56	&	1.02	&	$B-V$	\\
F571V1	&	79\phd		&	9.75	&	H0	&	\phn3	&	35	&	\phn83	&	\phn5	&	\phn9.21	&	\phn9.00	&	0.55	&	$B-V$	\\
F563V2	&	61\phd		&	7.05	&	H0	&	\phn3	&	29	&	111	&	\phn5	&	\phn9.51	&	\phn9.48	&	0.51	&	$B-V$	\\
F568V1	&	80\phd		&	9.55	&	H0	&	\phn3	&	40	&	124	&	\phn5	&	\phn9.53	&	\phn9.34	&	0.57	&	$B-V$	\\
F5631	&	45\phd		&	6.2\phn	&	H0	&	\phn3	&	11	&	111	&	\phn1	&	\phn9.59	&	\phn9.15	&	0.64	&	$B-V$	\\
UGC6446	&	18.6\phn	&	1.93	&	GROUP	&	\phn5	&	24	&	\phn83	&	\phn3	&	\phn9.64	&	\phn9.56	&	0.39	&	$B-V$	\\
UGC1230	&	51\phd		&	7\phd\phn\phn &	H0 	&	\phn3	&	16	&	133	&	\phn2	&	\phn9.91	&	\phn9.51	&	0.54	&	$B-V$	\\
UGC128	&	60\phd		&	8.4\phn	&	H0 	&	\phn3	&	21	&	130	&	\phn2	&	\phn9.96	&	\phn9.72	&	0.60	&	$B-V$	\\

\enddata
\tablecomments{Column (1) gives the galaxy's name; Column (2) gives the distance used for this paper; Column (3) gives the uncertainty for the distance; Column (4) gives the method used to estimate the distance; Column (5) gives the corresponding reference to the distance value; Column (6) gives the inclination of the galaxy; Column (7) gives the flat rotational velocity of the galaxy; Column (8) gives the uncertainty to the velocity measurement; Column (9) gives gas mass of the galaxy; Column (10) gives the B band luminosity of the galaxy; Column (11) gives the color measurement; Column (12) gives the band corresponding to each color measurement.}
\tablenotetext{a}{Distance estimation methods: $H_0$: Hubble flow with Virgo Cluster effect; CEPH: cepheid; TRGB: tip of red giant branch; GROUP: average distance of galaxy's group; BS: brightest stars; AVG: average of several measurements} 
\tablenotetext{b}{Inclination estimated with equation (1)}
\tablerefs{(1) \cite{sanders96}; (2) \cite{begeman91}; (3) \cite{deblok98}; (4) \cite{karachentsev05}; (5) \cite{tully00};(6) \cite{kassin06}; (7) \cite{matthews08}; (8) \cite{begum05}; (9) \cite{karachentsev04}; (10) \cite{jackson04}; (11) \cite{swaters99} ; (12) \cite{uson03}; (13) \cite{deblok01}; (14) \cite{begum03}; (15) \cite{warren04}
}
\end{deluxetable}

%% file: t2.tex
\begin{deluxetable}{lccccccccccccccc}
\rotate
\tabletypesize{\scriptsize}
\tablewidth{48pc}
\tablecaption{Mass Results from Portinari's Population Models}
\tablehead{
\colhead{}    & \colhead{}    & \multicolumn{4}{c}{Kroupa IMF} &   \colhead{}   &\multicolumn{4}{c}{Salpeter IMF}&   \colhead{}   & \multicolumn{4}{c}{Kennicutt IMF} \\
\cline{3-6} \cline{8-11} \cline{13-16}\\
\colhead{Galaxy} & \colhead{} & \colhead{$\Upsilon$}   & \colhead{$M_b$}    & \colhead{$\sigma_{M_b}$} & \colhead{$M_g>M_s$?} & \colhead{} &\colhead{$\Upsilon$}    & \colhead{$M_b$}   & \colhead{$\sigma_{M_b}$}   & \colhead{$M_g>M_s$?} & \colhead{} & \colhead{$\Upsilon$} & \colhead{$M_b$}& \colhead{$\sigma_{M_b}$} & \colhead{$M_g>M_s$?}\\
\colhead{} & \colhead{} & \colhead{($M_{\sun}/{L_{\sun}}$)} & \multicolumn{2}{c}{($\log{M_{\sun}}$)} & \colhead{} &\colhead{} & \colhead{($M_{\sun}/{L_{\sun}}$)} & \multicolumn{2}{c}{($\log{M_{\sun}}$)} & \colhead{} & \colhead{} & \colhead{($M_{\sun}/{L_{\sun}}$)} & \multicolumn{2}{c}{($\log{M_{\sun}}$)} & \colhead{}
}
\startdata
DDO210	&	&	0.31	&	6.61	&	0.13	&	Y	&	&	0.52	&	6.66	&	0.12	&	Y	&	&	0.26	&	6.59	&	0.14	&	Y	\\
CamB	&	&	1.14	&	7.41	&	0.12	&	Y	&	&	1.89	&	7.51	&	0.12	&	Y	&	&	0.92	&	7.38	&	0.12	&	Y	\\
WLM	&	&	0.37	&	7.81	&	0.12	&	Y	&	&	0.61	&	7.90	&	0.12	&	Y	&	&	0.30	&	7.78	&	0.12	&	Y	\\
NGC3741	&	&	0.52	&	8.36	&	0.15	&	Y	&	&	0.87	&	8.38	&	0.14	&	Y	&	&	0.42	&	8.36	&	0.15	&	Y	\\
UGC8550	&	&	2.32	&	8.60	&	0.19	&	Y	&	&	3.77	&	8.71	&	0.19	&	N	&	&	1.86	&	8.56	&	0.19	&	Y	\\
DDO168	&	&	0.41	&	8.73	&	0.13	&	Y	&	&	0.69	&	8.79	&	0.12	&	Y	&	&	0.33	&	8.71	&	0.13	&	Y	\\
NGC3109	&	&	0.74	&	8.70	&	0.15	&	Y	&	&	1.23	&	8.71	&	0.14	&	Y	&	&	0.60	&	8.69	&	0.15	&	Y	\\
UGC8490	&	&	0.28	&	8.96	&	0.13	&	Y	&	&	0.46	&	9.03	&	0.12	&	Y	&	&	0.23	&	8.94	&	0.13	&	Y	\\
UGC3711	&	&	1.32	&	9.21	&	0.12	&	N	&	&	2.14	&	9.33	&	0.13	&	N	&	&	1.06	&	9.16	&	0.12	&	Y	\\
F565V2	&	&	0.86	&	9.02	&	0.15	&	Y	&	&	1.44	&	9.07	&	0.14	&	Y	&	&	0.70	&	9.00	&	0.15	&	Y	\\
UGC5721	&	&	0.42	&	9.00	&	0.22	&	Y	&	&	0.69	&	9.04	&	0.21	&	Y	&	&	0.34	&	8.99	&	0.22	&	Y	\\
UGC3851	&	&	0.83	&	9.08	&	0.13	&	Y	&	&	1.35	&	9.14	&	0.12	&	Y	&	&	0.67	&	9.06	&	0.13	&	Y	\\
UGC6923	&	&	0.61	&	9.49	&	0.13	&	N	&	&	1.02	&	9.64	&	0.15	&	N	&	&	0.49	&	9.43	&	0.13	&	N	\\
IC2574	&	&	0.61	&	9.32	&	0.12	&	Y	&	&	1.02	&	9.43	&	0.12	&	N	&	&	0.49	&	9.28	&	0.12	&	Y	\\
UGC9211	&	&	0.54	&	9.13	&	0.24	&	Y	&	&	0.88	&	9.16	&	0.23	&	Y	&	&	0.43	&	9.13	&	0.24	&	Y	\\
NGC1560	&	&	1.09	&	9.26	&	0.12	&	Y	&	&	1.82	&	9.33	&	0.12	&	Y	&	&	0.89	&	9.23	&	0.13	&	Y	\\
NGC5585	&	&	0.71	&	9.37	&	0.19	&	Y	&	&	1.19	&	9.48	&	0.19	&	N	&	&	0.58	&	9.34	&	0.19	&	Y	\\
UGC7321	&	&	1.06	&	9.41	&	0.21	&	Y	&	&	1.73	&	9.51	&	0.22	&	N	&	&	0.85	&	9.37	&	0.21	&	Y	\\
UGC6818	&	&	0.63	&	9.49	&	0.12	&	N	&	&	1.06	&	9.62	&	0.14	&	N	&	&	0.51	&	9.44	&	0.12	&	Y	\\
UGC4499	&	&	0.54	&	9.33	&	0.16	&	Y	&	&	0.88	&	9.40	&	0.15	&	Y	&	&	0.43	&	9.31	&	0.17	&	Y	\\
IC 2233	&	&	0.47	&	9.40	&	0.12	&	Y	&	&	0.77	&	9.49	&	0.12	&	Y	&	&	0.38	&	9.37	&	0.12	&	Y	\\
F583-1	&	&	0.54	&	9.30	&	0.17	&	Y	&	&	0.90	&	9.33	&	0.16	&	Y	&	&	0.44	&	9.29	&	0.17	&	Y	\\
NGC6503	&	&	1.09	&	9.78	&	0.14	&	N	&	&	1.82	&	9.95	&	0.15	&	N	&	&	0.89	&	9.72	&	0.13	&	N	\\
UGC6917	&	&	0.93	&	9.90	&	0.13	&	N	&	&	1.56	&	10.06	&	0.15	&	N	&	&	0.76	&	9.85	&	0.13	&	N	\\
NGC4010	&	&	0.97	&	10.11	&	0.14	&	N	&	&	1.62	&	10.27	&	0.15	&	N	&	&	0.79	&	10.05	&	0.13	&	N	\\
UGC6983	&	&	0.68	&	9.87	&	0.12	&	Y	&	&	1.14	&	9.98	&	0.13	&	N	&	&	0.56	&	9.83	&	0.12	&	Y	\\
NGC2403	&	&	0.54	&	9.94	&	0.12	&	Y	&	&	0.90	&	10.06	&	0.13	&	N	&	&	0.44	&	9.89	&	0.12	&	Y	\\
NGC4183	&	&	0.54	&	10.07	&	0.13	&	N	&	&	0.90	&	10.22	&	0.14	&	N	&	&	0.44	&	10.02	&	0.13	&	N	\\
F5741	&	&	1.34	&	10.00	&	0.11	&	N	&	&	2.18	&	10.12	&	0.12	&	N	&	&	1.08	&	9.95	&	0.11	&	Y	\\
F5681	&	&	1.14	&	9.94	&	0.13	&	Y	&	&	1.89	&	10.04	&	0.13	&	Y	&	&	0.92	&	9.91	&	0.13	&	Y	\\
NGC3769	&	&	1.43	&	10.34	&	0.13	&	N	&	&	2.39	&	10.49	&	0.15	&	N	&	&	1.16	&	10.28	&	0.13	&	N	\\
NGC1003	&	&	1.01	&	10.40	&	0.21	&	N	&	&	1.68	&	10.56	&	0.23	&	N	&	&	0.82	&	10.35	&	0.21	&	N	\\
NGC5033	&	&	1.01	&	10.49	&	0.21	&	N	&	&	1.68	&	10.65	&	0.23	&	N	&	&	0.82	&	10.43	&	0.21	&	N	\\
NGC3198	&	&	0.63	&	10.45	&	0.12	&	Y	&	&	1.06	&	10.57	&	0.13	&	N	&	&	0.51	&	10.41	&	0.12	&	Y	\\
NGC2998	&	&	0.68	&	10.96	&	0.14	&	N	&	&	1.14	&	11.12	&	0.15	&	N	&	&	0.56	&	10.90	&	0.13	&	N	\\
NGC6674	&	&	1.09	&	11.05	&	0.15	&	N	&	&	1.82	&	11.21	&	0.17	&	N	&	&	0.89	&	11.00	&	0.15	&	N	\\

\enddata
\tablecomments{Not shown here are galaxies that are star dominated for all population models.  They are not used to determine the BTF relation.}
\end{deluxetable}



%% file: t3_corr.tex
\begin{deluxetable}{lccccccccccccccc}
\tabletypesize{\scriptsize}
\rotate
\tablewidth{48pc}
\tablecaption{Mass Results From Bell's Population Models}
\tablehead{
\colhead{}    & \colhead{}    & \multicolumn{4}{c}{Scaled-Salpeter IMF} &   \colhead{}   &
\multicolumn{4}{c}{Kroupa IMF}&   \colhead{}   &
\multicolumn{4}{c}{Bottema IMF} \\
\cline{3-6} \cline{8-11} \cline{13-16}\\
\colhead{Galaxy} & \colhead{} & \colhead{$\Upsilon$}   & \colhead{$M_b$}    & \colhead{$\sigma_{M_b}$} & \colhead{$M_g>M_s$?} & \colhead{} &\colhead{$\Upsilon$}    & \colhead{$M_b$}   & \colhead{$\sigma_{M_b}$}   & \colhead{$M_g>M_s$?} & \colhead{} & \colhead{$\Upsilon$} & \colhead{$M_b$}& \colhead{$\sigma_{M_b}$} & \colhead{$M_g>M_s$?}\\
\colhead{} & \colhead{} & \colhead{($M_{\sun}/{L_{\sun}}$)} & \multicolumn{2}{c}{($\log{M_{\sun}}$)} & \colhead{} &\colhead{} & \colhead{($M_{\sun}/{L_{\sun}}$)} & \multicolumn{2}{c}{($\log{M_{\sun}}$)} & \colhead{} & \colhead{} & \colhead{($M_{\sun}/{L_{\sun}}$)} & \multicolumn{2}{c}{($\log{M_{\sun}}$)} & \colhead{}
}
\startdata
DDO210	&	&	0.31	&	6.61	&	0.13	&	Y	&	&	0.22	&	6.58	&	0.14	&	Y	&	&	0.12	&	6.56	&	0.14	&	Y	\\
CamB	&	&	1.16	&	7.42	&	0.11	&	Y	&	&	0.82	&	7.37	&	0.12	&	Y	&	&	0.46	&	7.31	&	0.13	&	Y	\\
WLM	&	&	0.36	&	7.81	&	0.11	&	Y	&	&	0.26	&	7.76	&	0.12	&	Y	&	&	0.15	&	7.71	&	0.13	&	Y	\\
NGC3741	&	&	0.52	&	8.36	&	0.15	&	Y	&	&	0.37	&	8.35	&	0.15	&	Y	&	&	0.21	&	8.34	&	0.15	&	Y	\\
UGC8550	&	&	2.65	&	8.63	&	0.18	&	Y	&	&	1.88	&	8.56	&	0.18	&	Y	&	&	1.06	&	8.48	&	0.20	&	Y	\\
DDO168	&	&	0.41	&	8.73	&	0.13	&	Y	&	&	0.29	&	8.70	&	0.13	&	Y	&	&	0.16	&	8.66	&	0.14	&	Y	\\
NGC3109	&	&	0.75	&	8.70	&	0.15	&	Y	&	&	0.53	&	8.69	&	0.15	&	Y	&	&	0.30	&	8.68	&	0.15	&	Y	\\
UGC8490	&	&	0.36	&	8.99	&	0.12	&	Y	&	&	0.26	&	8.95	&	0.12	&	Y	&	&	0.14	&	8.91	&	0.14	&	Y	\\
UGC3711	&	&	1.55	&	9.24	&	0.10	&	N	&	&	1.10	&	9.17	&	0.11	&	Y	&	&	0.62	&	9.07	&	0.11	&	Y	\\
F565V2	&	&	0.88	&	9.02	&	0.15	&	Y	&	&	0.62	&	8.99	&	0.15	&	Y	&	&	0.35	&	8.96	&	0.16	&	Y	\\
UGC5721	&	&	0.53	&	9.02	&	0.21	&	Y	&	&	0.37	&	9.00	&	0.22	&	Y	&	&	0.21	&	8.97	&	0.23	&	Y	\\
UGC3851	&	&	1.00	&	9.10	&	0.12	&	Y	&	&	0.71	&	9.07	&	0.13	&	Y	&	&	0.40	&	9.03	&	0.14	&	Y	\\
UGC6923	&	&	0.61	&	9.49	&	0.11	&	N	&	&	0.43	&	9.40	&	0.11	&	N	&	&	0.24	&	9.28	&	0.11	&	Y	\\
IC2574	&	&	0.61	&	9.32	&	0.11	&	Y	&	&	0.43	&	9.26	&	0.11	&	Y	&	&	0.24	&	9.19	&	0.13	&	Y	\\
UGC9211	&	&	0.67	&	9.14	&	0.23	&	Y	&	&	0.47	&	9.13	&	0.24	&	Y	&	&	0.27	&	9.11	&	0.25	&	Y	\\
NGC1560	&	&	1.12	&	9.26	&	0.12	&	Y	&	&	0.79	&	9.22	&	0.13	&	Y	&	&	0.44	&	9.18	&	0.14	&	Y	\\
NGC5585	&	&	0.72	&	9.38	&	0.18	&	Y	&	&	0.51	&	9.32	&	0.19	&	Y	&	&	0.29	&	9.25	&	0.21	&	Y	\\
UGC7321	&	&	1.26	&	9.44	&	0.20	&	Y	&	&	0.89	&	9.38	&	0.21	&	Y	&	&	0.50	&	9.29	&	0.22	&	Y	\\
UGC6818	&	&	0.64	&	9.49	&	0.11	&	N	&	&	0.45	&	9.42	&	0.11	&	Y	&	&	0.25	&	9.32	&	0.12	&	Y	\\
UGC4499	&	&	0.67	&	9.36	&	0.15	&	Y	&	&	0.47	&	9.32	&	0.16	&	Y	&	&	0.27	&	9.27	&	0.18	&	Y	\\
IC 2233	&	&	0.59	&	9.44	&	0.11	&	Y	&	&	0.42	&	9.38	&	0.11	&	Y	&	&	0.23	&	9.31	&	0.13	&	Y	\\
F583-1	&	&	0.54	&	9.30	&	0.17	&	Y	&	&	0.39	&	9.28	&	0.18	&	Y	&	&	0.22	&	9.26	&	0.18	&	Y	\\
NGC6503	&	&	1.12	&	9.79	&	0.11	&	N	&	&	0.79	&	9.69	&	0.10	&	N	&	&	0.44	&	9.55	&	0.11	&	Y	\\
UGC6917	&	&	0.95	&	9.91	&	0.11	&	N	&	&	0.67	&	9.82	&	0.11	&	N	&	&	0.38	&	9.70	&	0.11	&	Y	\\
NGC4010	&	&	0.99	&	10.11	&	0.11	&	N	&	&	0.70	&	10.02	&	0.11	&	N	&	&	0.39	&	9.88	&	0.11	&	Y	\\
UGC6983	&	&	0.69	&	9.87	&	0.11	&	Y	&	&	0.49	&	9.80	&	0.11	&	Y	&	&	0.28	&	9.73	&	0.12	&	Y	\\
NGC2403	&	&	0.54	&	9.94	&	0.11	&	Y	&	&	0.39	&	9.87	&	0.11	&	Y	&	&	0.22	&	9.79	&	0.12	&	Y	\\
NGC4183	&	&	0.54	&	10.08	&	0.11	&	N	&	&	0.39	&	9.99	&	0.11	&	N	&	&	0.22	&	9.89	&	0.11	&	Y	\\
F5741	&	&	1.59	&	10.04	&	0.09	&	N	&	&	1.12	&	9.96	&	0.10	&	Y	&	&	0.63	&	9.87	&	0.10	&	Y	\\
F5681	&	&	1.16	&	9.95	&	0.12	&	Y	&	&	0.82	&	9.90	&	0.13	&	Y	&	&	0.46	&	9.84	&	0.14	&	Y	\\
NGC3769	&	&	1.48	&	10.34	&	0.11	&	N	&	&	1.05	&	10.25	&	0.11	&	N	&	&	0.59	&	10.13	&	0.11	&	Y	\\
NGC1003	&	&	1.03	&	10.41	&	0.20	&	N	&	&	0.73	&	10.32	&	0.19	&	N	&	&	0.41	&	10.19	&	0.19	&	Y	\\
NGC5033	&	&	1.03	&	10.49	&	0.20	&	N	&	&	0.73	&	10.40	&	0.19	&	N	&	&	0.41	&	10.27	&	0.19	&	Y	\\
NGC3198	&	&	0.64	&	10.45	&	0.11	&	Y	&	&	0.45	&	10.38	&	0.11	&	Y	&	&	0.25	&	10.30	&	0.12	&	Y	\\
NGC2998	&	&	0.69	&	10.96	&	0.11	&	N	&	&	0.49	&	10.87	&	0.11	&	N	&	&	0.28	&	10.74	&	0.11	&	Y	\\
NGC6674	&	&	1.12	&	11.06	&	0.13	&	N	&	&	0.79	&	10.97	&	0.13	&	N	&	&	0.44	&	10.84	&	0.13	&	Y	\\

\enddata
\tablecomments{Not shown here are galaxies that are star dominated for all population models.  They are not used to determine the BTF relation.}
\end{deluxetable}

%% file: t4_corr.tex
\begin{deluxetable}{lcccccccccccccc}
\tablewidth{52pc}
\rotate
\tablecaption{BTF Fit to Gas Dominated Galaxies}
\tablehead{
\colhead{Subsample} & \colhead{N} & \colhead{$x_{v|M}$} & \colhead{$\sigma_{x_{v|M}}$} & \colhead{$A_{v|M}$} & \colhead{$\sigma_{A_{v|M}}$} & \colhead{$\chi_{\nu,v|M}^2$} & \colhead{$x_{M|v}$} & \colhead{$\sigma_{x_{M|v}}$} & \colhead{$A_{M|v}$} & \colhead{$\sigma_{A_{M|v}}$} & \colhead{$\chi_{\nu,M|v}^2$} & \colhead{$x_{bis}$} & \colhead{$A_{bis}$} &\colhead{$\chi_{\nu,bis}^2$}}

\startdata
Portinari-Kroupa &	23 &	3.77 &	0.22 &  2.08 &	0.42 &	1.28 &	4.11 &	0.25 &	1.43 &	0.47 &	1.18 &	3.93 &	1.78 & 1.25\\
Portinari-Salpeter &	14 &	3.59 &	0.40 &	2.44 &	0.73 &	1.42 &	4.37 &	0.50 &	1.02 &	0.91 &	1.46 &	3.94 &	1.79 & 1.46\\
Portinari-Kennicutt &	26 &	3.74 &	0.21 &	2.14 &	0.41 &	2.01 &	4.33 &	0.26 &	0.99 &	0.50 &	1.85 &	4.01 &	1.62 & 1.99\\
Bell-Scaled Salpeter &	23 &	3.77 &	0.20 &	2.09 &	0.39 &	1.41 &	4.09 &	0.23 &	1.47 &	0.45 &	1.31 &	3.93 &	1.80 & 1.37\\ 
Bell-Kroupa &		26 &	3.72 &	0.20 &  2.17 &	0.40 &  2.30 &	4.36 &	0.25 &  0.94 &	0.49 &  2.10 &	4.01 &	1.61 & 2.27\\
Bell-Bottema &		36 &	3.55 &	0.29 &  2.45 &  0.29 &  2.02 &	3.96 &	0.16 &  1.63 &	0.32 &  2.06 &	3.74 &	2.06 & 2.04\\

\enddata

\end{deluxetable}

%% file: t5_corr.tex
\begin{deluxetable}{lccc}
\tabletypesize{\footnotesize}
\tablewidth{0pt}
\tablecaption{Derived Values from the BTF Relation}
\tablehead{
\colhead{Galaxy} &  \colhead{$M_b$} & \colhead{$M_s$\tablenotemark{a}} 
& \colhead{$\Upsilon$} \\
\colhead{} & \multicolumn{2}{c}{($\log{M_{\sun}}$)} 
& \colhead{($M_{\sun}/{L_{\sun}}$)}
}
\startdata

KK98251        &        \phn7.92        &        \phn-7.50        &        -0.28        \\
NGC2683        &        10.42        &        \phs10.40        &       \phs 1.82        \\
UGC6667        &       \phn9.39        &        \phs\phn9.12        &       \phs 0.35        \\
UGC6923        &        \phn9.31        &        \phs\phn8.95        &     \phs   0.28        \\
UGC6818        &        \phn9.11        &        \phn-8.21        &        -0.06        \\
NGC7793        &        \phn9.58        &        \phs\phn9.35        &       \phs 0.42        \\
NGC3972        &        10.16        &        \phs10.10        &       \phs 1.29        \\
NGC4085        &        10.20        &        \phs10.14        &        \phs1.18        \\
NGC6503        &        \phn9.91        &       \phs\phn 9.79        &       \phs 1.63        \\
NGC3877        &        10.58        &        \phs10.55        &        \phs1.27        \\
NGC4138        &        10.34        &        \phs10.30        &       \phs 1.70        \\
NGC247        &        \phn9.77        &        \phs\phn9.57        &       \phs 0.63        \\
UGC6973        &        10.62        &        \phs10.59        &        \phs4.38        \\
NGC3917        &        10.21        &        \phs10.13        &        \phs0.84        \\
UGC6917        &        \phn9.85        &       \phs\phn 9.62        &      \phs  0.76        \\
NGC4217        &        10.66        &        \phs10.62        &       \phs 1.52        \\
NGC4051        &        10.47        &        \phs10.42        &       \phs 0.70        \\
NGC3953        &        11.04        &        \phs11.03        &       \phs 2.55        \\
NGC4010        &        10.02        &       \phs\phn 9.82        &        \phs0.73        \\
NGC4013        &        10.65        &        \phs10.60        &       \phs 1.92        \\
NGC4100        &        10.47        &        \phs10.41        &       \phs 0.99        \\
NGC3949        &        10.53        &        \phs10.46        &       \phs 1.20        \\
NGC4183        &        \phn9.85        &        \phs\phn9.33        &       \phs 0.17        \\
NGC2903        &        10.69        &        \phs10.63        &       \phs 1.44        \\
NGC3521        &        10.77        &        \phs10.71        &       \phs 1.97        \\
NGC3769        &        \phn9.95        &        \phs\phn9.14        &     \phs   0.14        \\
NGC3893        &        10.64        &        \phs10.55        &       \phs 1.14        \\
NGC1003        &        \phn9.86        &       \phn -9.20        &        -0.10        \\
NGC3726        &        10.56        &        \phs10.43        &       \phs 0.72        \\
NGC5033        &        10.81        &        \phs10.74        &        \phs2.69        \\
NGC5907        &        10.98        &        \phs10.93        &        \phs3.52        \\
NGC7331        &        11.15        &        \phs11.12        &       \phs 2.39        \\
NGC4088        &        10.58        &        \phs10.42        &        \phs0.65        \\
NGC4157        &        10.72        &        \phs10.62        &        \phs1.43        \\
NGC3992        &        11.18        &        \phs11.14        &        \phs3.06        \\
NGC2998        &        10.96        &        \phs10.78        &        \phs0.67        \\
NGC801        &        10.99        &        \phs10.82        &        \phs0.83        \\
NGC5533        &        11.17        &        \phs11.06        &        \phs1.91        \\
NGC2841        &        11.47        &        \phs11.41        &        \phs1.36        \\
NGC6674        &        11.18        &        \phs11.05        &        \phs1.67        \\
UGC2885        &        11.54        &        \phs11.47        &        \phs1.30        \\

\enddata

\tablenotetext{a}{Sometimes the baryonic mass from the BTF is less than the gas mass, 
leading to a negative stellar mass.  This should happen occassionally 
because of scatter in the relation.  In no case is the stellar mass 
significantly less than zero.
}
\end{deluxetable}


